\documentclass[11pt]{article}

\usepackage{graphicx}
\usepackage{bm}

\begin{document}
\begin{titlepage}

\begin{flushright}
CERN--PH--TH/2005--19
\end{flushright}

\vspace*{1cm}

\begin{center}
\Large{\bf  Scalar perturbations in  regular two-component
bouncing cosmologies}

\vspace*{1cm}

\large{ V. Bozza${}^{a,b,c}$ and G. Veneziano${}^{d,e}$}

\bigskip
\normalsize

{\sl $^a$ Centro studi e ricerche ``Enrico Fermi'', via Panisperna
89/A, Rome, Italy. \\
 $^b$ Dipartimento di Fisica ``E. R. Caianiello", Universit\`a di
Salerno, I-84081 Baronissi, Italy. \\
 $^c$ Istituto Nazionale di Fisica Nucleare, Sezione di Napoli,
 Naples, Italy.\\
 $^d$ CERN, PH Dept., TH Unit, CH-1211, Geneva 23, Switzerland.\\
 $^e$ Coll\`ege de France, 11 place Marcelin-Berthelot, 75005 Paris, France.}


\vspace*{5mm}

\begin{abstract}
We consider a two-component regular cosmology bouncing from
contraction to expansion, where, in order to include both scalar
fields and perfect fluids as particular cases, the dominant
component is allowed to have an intrinsic isocurvature mode. We
show that the spectrum of the growing mode of the Bardeen
potential in the pre-bounce is never transferred to the dominant
mode of the post-bounce. The latter acquires at most a dominant
isocurvature component, depending on the relative properties of
the two fluids. Our results imply that several claims in the
literature need substantial revision.
\end{abstract}

\end{center}


\end{titlepage}

Bouncing cosmologies have been proposed as possible alternatives
to standard inflation in string-inspired  (e.g. Pre-Big Bang
\cite{PBB}, ekpyrotic/cyclic \cite{Ekp}) scenarios. In order to
compare the predictions of these scenarios with observations, it
is crucial to follow the evolution of cosmological perturbations
from the initial vacuum-normalized state, through the bounce, all
the way until decoupling. Nevertheless, there is no general
agreement on the true influence of a bounce on cosmological
perturbations. In particular, it is not clear whether the
Pre-Bounce growing mode of the Bardeen potential leaves any trace
in the Post-Bounce, or whether  it just completely matches a
decaying mode \cite{Others}. Only in the former case can the claim
that the ekpyrotic/cyclic scenario is in agreement with CMB data
be defended.

Studies of specific regular models apparently led to different
results, suggesting that both possibilities can arise, depending
on the specific model one assumes for the regular bounce. In
spatially closed universes, a regular bounce from contraction to
expansion can be triggered by the spatial curvature term in the
Friedman equations \cite{K=1}. In this case, the transfer matrix
relating Post-Bounce modes to Pre-Bounce perturbations depends on
the momentum scale $k$, suggesting that non-trivial matching
conditions hold. These models have the advantage of being
completely embedded into General Relativity, but they need
fine-tuning in the initial conditions in order to have a
sufficiently long collapse.

Alternatively, one can start from a spatially flat universe,
avoiding the complications of spatial curvature. However, Friedman
equations imply that a bounce from contraction to expansion
necessarily violates the null energy condition (NEC). This
violation can be induced by high energy and high curvature
corrections to General Relativity or can be directly introduced by
a ghost field dominating during the bounce. In this class of
models, contradictory results have emerged up to now. Peter \&
Pinto-Neto \cite{PetPin} and Finelli \cite{Fin} find some mixing
between the growing and decay modes of the Bardeen potential,
while Gasperini, Giovannini \& Veneziano \cite{GGV}, Cartier
\cite{Cartier}, and Allen \& Wands \cite{AllWan} find that the
growing mode of the Bardeen potential in the Pre-Bounce exactly
matches the decaying mode in the Post-Bounce.

In this letter we report the results of the study of a  wide class
of regular bouncing cosmological models, containing,  as
particular cases, all the above-mentioned two-source regular
models. We confirm that the Pre-Bounce Bardeen growing mode
matches the Post-Bounce decaying mode. The only possible mode
mixing is between the two components that are present in the
cosmological description, as a consequence of the normal interplay
between adiabatic and isocurvature perturbations. We present
hereafter the key arguments and give the main results, postponing
most details to a later publication  \cite{BozVen2}.

We consider a two-component cosmology, where the first component
dominates at early and late times, while the secondary one has
negative energy density and triggers the bounce from contraction
to expansion in a fully regular fashion. It is well known that
ghosts or negative energy density fluids induce instabilities at a
quantum level, but this has nothing to do with the evolution of
classical perturbations, which is completely under control, as we
shall explicitly show. One may regard the secondary field as an
effective term in our 4-dimensional Friedman equations simulating
corrections from high energy, high curvature, extra-dimensions,
finite string size, or whatever else one can imagine. One can then look
at the regular bounce obtained in such an effective description at a
classical level using the paradigm of General Relativity, without
worrying about quantum instabilities which should be properly
addressed only within a yet unavailable complete theory of the bounce. These
considerations confer a huge interest to ghost-induced-bounces as
the only effective models of spatially-flat bouncing cosmologies
which can be investigated within the frame and with the
instruments of General Relativity. In some sense, they are the
only models worth speaking about at the effective level!

The background equations are
\begin{eqnarray}
&& 3\mathcal{H}^{2}=a^2 \rho \\ &&
\mathcal{H}^2+2\mathcal{H}'=-a^2 p
\\ && \rho'+3\mathcal{H}(\rho+p)=0, \label{Eqrho'}
\end{eqnarray}
where a prime denotes derivative w.r.t. conformal time $\eta$, $a$
is the scale factor, $\mathcal{H}=a'/a$, and we have set $8\pi
G=1$. The total energy density is $\rho=\rho_a+\rho_b$ and the
pressure is $p=p_a+p_b$. The two fluids satisfy the continuity
equation (\ref{Eqrho'}) independently, i.e. we assume no
interaction between the two fluids other than gravitational. We
shall set $\eta=0$ at the bounce, i.e. when $\mathcal{H}$ changes
sign. A typical background solution starts from a contraction
driven by the first fluid, which ends when the secondary fluid's
density becomes of the same order as that of the first fluid. We
will call this time $-\eta_b$ and normalize $\eta$ in such a way
that $\eta_b \simeq 1$. As we are restricting ourselves to
spatially-flat cosmologies, the null energy condition (NEC,
hereafter) is necessarily violated, i.e. $\rho+p$ must change sign
at some time $\eta_{\mathrm{NEC}}$, with
$-1<\eta_{\mathrm{NEC}}<0$. After the bounce the NEC is restored,
the secondary fluid becomes subdominant again, and an expansion
phase dominated by the first fluid follows.

We shall consider only scalar perturbations, since these are both
the most relevant and the most controversial. The perturbed line
element takes the well-known form
\begin{eqnarray}
&ds^2=&a^2(\eta)\left\{ (1+2\phi) d\eta^2-2B_{,i} d\eta dx^i
\right. \nonumber \\ && \left.
-\left[(1-2\psi)\delta_{ij}+2E_{,ij}\right]dx^i dx^j \right\},
\end{eqnarray}
while the perturbed energy--momentum tensor of the first source
reads:
\begin{equation}
{{(T_a)}^\mu}_\nu= \left(%
\begin{array}{cc}
  \rho_a+\delta \rho_a & -(\rho_a+p_a) \mathcal{V}_{a,i} \\
  (\rho_a+p_a) \mathcal{V}_{a,i} & -(p_a+\delta p_a) \delta_{ij} \;  \\
\end{array}%
\right), \label{TmunuP}
\end{equation}
where $\mathcal{V}_a$ is its velocity potential. The same
expression holds for the secondary fluid, the subscript $a$ being
replaced by $b$.

In order to keep our treatment as general as possible, we
introduce the following gauge-invariant relations among the
perturbations of the sources:
\begin{equation}
\!\!\!\! \delta p_a = c_a^2 \delta \rho_a + \alpha (\rho_a + p_a)
\mathcal{V}_a; \;\; \alpha = \frac{p_a'-c_a^2
\rho_a'}{\rho_a+p_a}; \;\; \delta p_b = c_b^2 \delta \rho_b .
\end{equation}
Notice that if $c_a^2 \neq p_a'/\rho_a'$, the additional
dependence on the velocity potential generates an isocurvature
mode
\begin{equation}
\tau_a \delta S_a = \delta p_a- (p_a'/\rho_a')\delta
\rho_a=\alpha(\rho_a + p_a)\left(\mathcal{V}_a-\delta
\rho_a/\rho_a'\right).
\end{equation}
It is a trivial exercise to prove that a scalar field $\varphi$
with a self-interaction potential $V(\varphi)$ satisfies these
relations with $c_a^2=1$ and $\alpha=-2a^2V_{,\varphi}/\varphi'$.
Therefore, these relations include perfect fluids and scalar
fields as particular cases of a general class of sources and
ensure a wide generality to our treatment. Allen \& Wands' model
\cite{AllWan} is recovered with the scalar field choice for the
first fluid and $c_b^2=1$. The perfect-fluid model studied by
Finelli  \cite{Fin} is recovered setting $\alpha=0$,
$c_b^2=2c_a^2+1/3$. The model by Peter \& Pinto-Neto \cite{PetPin}
corresponds to $\alpha=0$, $c_a^2=1/3$, $c_b^2=1$. The
generalization to the case where an intrinsic isocurvature mode
for the secondary fluid is present poses no major difficulties,
but will not be considered here for simplicity.

The first subtlety in the study of perturbations in cosmological
bounces is the choice of a gauge where all variables stay finite
and small enough for the linear theory to apply.  In particular,
we have to be sure that the perturbative variables we are
considering will not diverge at any point. From the physical point
of view the requirement is that all components of the metric and
of the energy--momentum tensor stay finite at least in one
convenient  gauge. If we assume that such a gauge exists (if it
does not  then the model under consideration cannot be a stable
solution of Einstein equations), we deduce that $\phi$, $\psi$,
$B$, $E$, $\delta \rho_i$, $\delta p_i$ and
$(\rho_i+p_i)\mathcal{V}_i$ stay finite and small in that gauge.
We will refer to it as the regular gauge. Note that the total
velocity potential $\mathcal{V}$ is allowed to diverge in the
regular gauge at  $\eta = \eta_{\mathrm{NEC}}$ when $\rho+p=0$
(NEC violation), since what matters is that the energy--momentum
tensor components stay finite. Our attitude will be just to assume
that such a regular gauge exists and to build gauge-invariant
quantities out of perturbations in this gauge.

The Bardeen potential $\Psi=\psi+\mathcal{H}(E'-B)$ is indeed a
first regular combination. The widely used
$\zeta=\psi+\mathcal{H}\mathcal{V}$ may diverge at
$\eta_{\mathrm{NEC}}$. However, we can replace it by
\begin{equation}
\tilde{\zeta}=(\mathcal{H}^2-\mathcal{H}')(\psi+\mathcal{H}\mathcal{V}),
\end{equation}
which is guaranteed to be regular, since
$2(\mathcal{H}^2-\mathcal{H}')\mathcal{V}=a^2(\rho+p)\mathcal{V}$
is finite in the regular gauge.  We then introduce individual
gauge-invariant variables
\begin{eqnarray}
&&\tilde{\zeta}_a= \frac{1}{2}a^2(\rho_a+p_a)
(\psi+\mathcal{H}\mathcal{V}_a ) \\ && \Psi_a= \frac{1}{2} a^2
(\rho_a' \psi+ \mathcal{H}\delta \rho_a ),
\end{eqnarray}
and similarly for the $b$ fluid. All these variables are regular
by construction. From the definition of $\tilde{\zeta}$ and the 00
component of the Einstein equations, we have
\begin{equation}
\tilde{\zeta}=\tilde{\zeta}_a+\tilde{\zeta}_b; \;\; \;\;
(3\tilde{\zeta}+k^2 \Psi)\mathcal{H}+ \Psi_a+\Psi_b = 0.
\end{equation}

The perturbations of the continuity equations of each fluid can be
combined to give coupled second order equations for
$\tilde{\zeta}_a$ and $\tilde{\zeta}_b$. Introducing the canonical
(Sasaki--Mukhanov) variables \cite{MV}  $v_{a}=
\tilde{\zeta}_a/[c_a (\rho_a+p_a)^{1/2}\mathcal{H}]$ and $v_{b}=
\tilde{\zeta}_b/[c_b (\rho_b+p_b)^{1/2}\mathcal{H}]$, we find
\begin{eqnarray}
&& v_{a}''+\left(c_a^2k^2-z_a''/z_a \right)v_{a} =
O(\rho_b/\rho_a)^{1/2} \label{vMa}\\ &&
v_{b}''+\left(c_b^2k^2-z_b''/z_b \right)v_{b} =
O(\rho_b/\rho_a)^{1/2}, \label{vMb}
\end{eqnarray}
with $z_a=a^2(\rho_a+ p_a)^{1/2}/(c_a \mathcal{H})$ and
$z_b=a^{(1+3c_b^2)}(\rho_b+ p_b)^{1/2}/c_b$. These equations also
identify $c_a^2$ and $c_b^2$ as the correct definitions of the
sound speed for the two components.

These equations decouple only in the limit $\rho_b/\rho_a
\rightarrow 0$. Both Mukhanov variables are normalized to vacuum
fluctuations and thus are of order $1/\sqrt{2k}$ in the asymptotic
past. However, we have
$\tilde{\zeta}_b=\sqrt{(\rho_b+p_b)/(\rho_a+p_a)}\tilde{\zeta}_a$
and a hierarchy can therefore be established between these two
variables in this limit. In this sense, the
$O(\rho_b/\rho_a)^{1/2}$ contains all  terms that are irrelevant
in the asymptotic past, including couplings between the two
variables.

Although it is not necessary, let us assume for simplicity that
the Pre-Bounce can be described by a power law $a(\eta) \simeq
(-\eta)^q$. Defining $\Gamma=\frac{p'}{\rho'}$, the following
general relations are implied by the background equations
\begin{equation}
q=\frac{2}{1+3\Gamma}, \; \; \alpha \sim
\frac{q(1+3c_a^2)-2}{-\eta}. \label{qGamma}
\end{equation}
If $\alpha=0$, we get $\Gamma=c_a^2$, which is the case of a
perfect fluid. A well-known case of a source where $\alpha$
contributes to the background is the case of a scalar field with
an exponential potential $e^{-\lambda \varphi}$. Then $c_a^2=1$,
but $\Gamma$ can assume any value depending on $\lambda$. This
case is discussed in Ref. \cite{AllWan}. A stiffer
$\eta$-dependence for $\alpha$ would make it rapidly subdominant,
while a softer one would render the vacuum normalization
problematic and will not be considered.

The secondary field evolves according to its continuity equation
$\rho_b \sim (-\eta)^{-3q(1+c_b^2)}$. Exploiting the fact that the
two fluids have comparable densities at time $\eta=-1$, at earlier
times we find $\rho_b/\rho_a \simeq (-\eta)^{-q(1+3c_b^2)+2}$.

In the power law regime, Eqs. (\ref{vMa}) and (\ref{vMb}) give
\begin{eqnarray}
&  v_{a}=& \sqrt{|\eta|}C_a H^{(1)}_{\nu_a}(c_a \sqrt{k} |\eta|)
\label{PowLawa}\\ &  v_{b}=& \sqrt{|\eta|} C_b H^{(1)}_{\nu_b}(c_b
\sqrt{k} |\eta|).\label{PowLawb}
\end{eqnarray}
with $\nu_a=\frac{1}{2}-q$, $\nu_b=\frac{1}{2}(3c_b^2q-1-q)$ and
$H_\nu^{(1)}$ is the Hankel function of the first kind; $C_a$ and
$C_b$ are pure numbers of order 1. The initial conditions for
$\tilde{\zeta}_a$, $\tilde{\zeta}_b$, $\Psi_a$ and $\Psi_b$ can be
derived from those we just imposed on  Mukhanov's variables.

From the perturbations of the continuity equations, it is
straightforward to write a set of first-order differential
equations for $\tilde{\zeta}$, $\Psi$, $\tilde{\zeta}_b$ $\Psi_b$.
However, it is more useful for our purposes to write these
equations in their integral form
\begin{eqnarray}
& \!\!\!\!\!\!\!\! \Psi & \!\!\!\! =
 \frac{\mathcal{H}}{a^2}
\left[\frac{c_1(k)}{k^2}+ \int \frac{a^2}{\mathcal{H}^2}
\tilde{\zeta} d\eta \right] \label{Psiint}
\\
& \!\!\!\!\!\!\!\! \tilde{\zeta} & \!\!\!\! = a^2(\rho_a+p_a)
\left[ c_2(k) -\int \frac{c_a^2\mathcal{H}}{a^2(\rho_a+p_a)} k^2
\Psi d\eta \right. \nonumber \\ && \left. -\int
\frac{\alpha}{a^2(\rho_a+p_a)} \tilde{\zeta}_b d\eta + \int
\frac{c_b^2-c_a^2}{a^2(\rho_a+p_a)}\Psi_b d\eta \right]
\label{zetarint}
\\
& \!\!\!\!\!\!\!\! \Psi_b & \!\!\!\! =
\frac{\mathcal{H}}{a^{1+3c_b^2}} \left[ c_3(k) -\int
\frac{a^{1+3c_b^2}k^2}{\mathcal{H}} \tilde{\zeta}_b d\eta \right.
\nonumber \\ && \left.  + \int \frac{a^{3+3c_b^2}(\rho_b +
p_b)}{2\mathcal{H}}k^2 \Psi d\eta \right] \label{Psibint}\\ &
\!\!\!\!\!\!\!\!\tilde{\zeta}_b & \!\!\!\! =
a^{1+3c_b^2}\mathcal{H} (\rho_b +p_b) \left[ c_4(k)+\int
\frac{c_b^2}{a^{1+3c_b^2}\mathcal{H} (\rho_b +p_b)} \Psi_b d\eta
\right. \nonumber \\ && \left. + \int
\frac{a^{1-3c_b^2}}{2\mathcal{H}^2}
 \tilde{\zeta} d\eta \right] \, , \label{zbint}
\end{eqnarray}

\begin{figure}
\resizebox{\hsize}{!}{\includegraphics{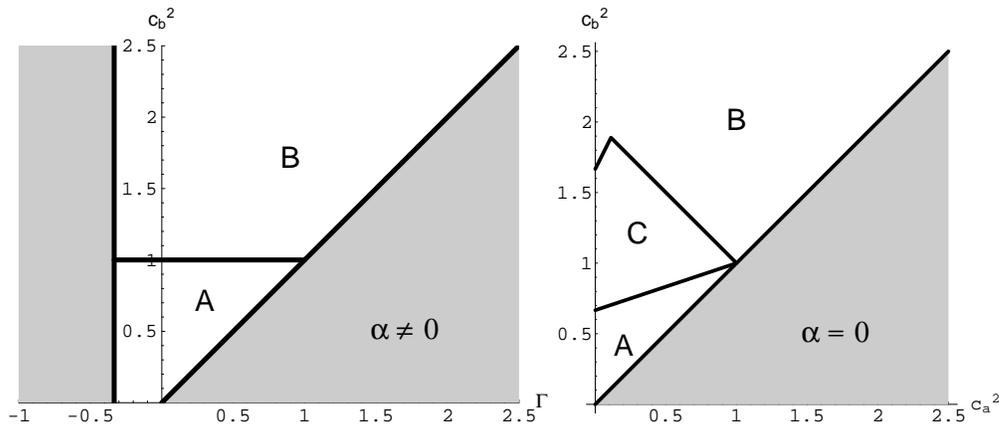}} \caption{ The
parameter space in the general case (left) and in the perfect
fluid case (right).}
 \label{Fig}
\end{figure}

Now we are ready to analyse the evolution of the modes that are
well outside the horizon at the bounce ($k \ll 1$). The discussion
can be made in terms of the two parameters $\Gamma$ and $c_b^2$,
which express the background evolution of the two fluids. Not all
values of these two parameters are relevant to bouncing
cosmologies. Indeed, as a first constraint, the condition that
$\rho_b/\rho_a$ must grow during contraction (otherwise there is
no bounce), excludes the half-plane $c_b^2<\Gamma$. A correct
vacuum normalization requires $c_a^2,c_b^2>0$. Finally, we require
$\Gamma>-1/3$, in order to discard superinflationary backgrounds.
For the sake of generality we do not impose the causality constraint
$c_a^2,c_b^2 \leq 1$. In Fig. \ref{Fig}, we show the allowed
regions in the cases $\alpha \neq 0$ and $\alpha=0$, the forbidden
ones being shaded.

The  four constants $c_i(k)$ can be determined by matching them to
the long-wavelength limit of Eqs. (\ref{PowLawa}),(\ref{PowLawb}):
\begin{equation}
c_1(k) \sim k^{\nu_a}, \; c_2(k) \sim  k^{-\nu_a}, \; c_3(k) \sim
k^{-\nu_b}, \; c_4(k) \sim  k^{\nu_b}.
\end{equation}

Now we have to compute the evolution of the individual terms in
Eqs. (\ref{Psiint})--(\ref{zbint}) and find the dominant
contribution  at horizon re-entry after the bounce. We will
discard any numerical factors of order 1, just keeping the
dependences on $k$ and $\eta$, the only dimensional parameters in
our analysis. In these equations, each variable is expressed as
the sum of a ``homogeneous" mode and several integrals, one for
each coupling term. The homogeneous mode of each variable persists
through the bounce and after the bounce, and thus poses no
difficulty. To evaluate the integrals, we note that, far from the
bounce, we can replace all the functions by their asymptotic power
law  behaviours . Thus each integrand $f(\eta)$ assumes the
generic form $c_i(k) |\eta|^s$ and, for each integral, we
generically have, in the Pre-Bounce:
\begin{equation}
\int\limits^\eta c_i(k)|\eta|^s d\eta \sim c_i(k) |\eta|^{s+1},
\;\;\; \eta<-1, \label{IntPreBounce}
\end{equation}
where the integration constant is set to zero in order to match,
in the far past, the asymptotic solutions implied by
(\ref{PowLawa}), (\ref{PowLawb}).

To evaluate the integrals in the Post-Bounce, we can split the
integration domain in three intervals, covering the full
Pre-Bounce, the Bounce and the Post-Bounce separately. We have
\begin{eqnarray}
&& \int\limits^{-1}_{} c_i(k)|\eta|^s d\eta  \sim  c_i(k), \;\;
\int\limits_{-1}^{1} f(\eta) d\eta \sim c_i(k), \nonumber \\&&
 \int\limits_{1}^{\eta} c_i(k)|\eta|^s d\eta \sim
c_i(k)\left[1-\eta^{s+1} \right].
\end{eqnarray}
The first integral is just Eq. (\ref{IntPreBounce}) evaluated at
$\eta=-1$, while the third  is evaluated using the power-law
expansion for the integrand $f(\eta)$ in the Post-Bounce. More
subtle is the bounce contribution. Since all the curvature and
time scales are fixed by the time normalization choice $\eta_b=1$,
the only scales that can be introduced by this contribution are
$k$ and $\eta$. Yet, the integrand is a combination of
$k$-independent backgrounds, and  integration over the finite
bounce interval eliminates $\eta$ from the final result. Hence the
bounce contribution is of the same order as $c_i(k)$. An apparent
complication comes from the divergence of some integrands in Eqs.
(\ref{Psiint})--(\ref{zbint}) at the bounce. Of course, these
divergences do not affect the regularity of the gauge-invariant
variables, since all the integrals are multiplied by proper
factors, which compensate these divergences. Mathematically we can
further split the bounce integrals in two pieces, before and after
$\eta=0$. The limiting value of the gauge-invariant variable at
$0^-$ fixes the initial condition for the second integral starting
from $0^+$ (recall that our integral equations are just another
representation of differential equations, which admit a regular
expansion for the solution at $\eta=0$). Afterwards, the
dimensional argument can be applied safely.

At late times ($\eta \gg 1$) the dominant term depends on the
value of $s$
\begin{equation}
\int\limits^{\eta}f(\eta)d\eta \sim \left\{
\begin{array}{ll}
c_i(k), &  s<-1 \\ c_i(k)\eta^{s+1}, & s>-1.
\end{array}
\right.
\end{equation}
Calculating all the integrals recursively, we can build a
$k$-expansion for each of the four independent modes $c_i(k)$.
Evaluating the solutions in this way obtained at the horizon
re-entry $\eta \sim 1/k$, we can pick out the dominant
contributions and read their spectra.

Let us first discuss the general case where $\alpha$ is different
from zero. This includes bounces with tracking scalar fields and
all asymmetric bounces. We can distinguish two regions in the
plane ($\Gamma$,$c_b^2$), illustrated in the left panel of Fig.
\ref{Fig}. In region A (defined by $c_b^2<1$), $\Psi$ and
$\tilde{\zeta}$ develop a typical adiabatic spectrum, determined
by the dominant fluid. We have $\Psi^A \sim \tilde{\zeta}^A \sim
c_1(k)$. In region B, covering the rest of the plane, $\Psi$ and
$\tilde{\zeta}$ acquire the spectrum of the secondary component,
so that $\Psi^B \sim \tilde{\zeta}^B\sim c_3(k)$. The individual
variables $\tilde{\zeta}_a$ and $\tilde{\zeta}_b$ follow the
behaviour of $\tilde{\zeta}$. Notice that on the border line
($c_b^2=1$), the two spectra coincide. It was this line that was
explored in Ref. \cite{AllWan}. In the particular case $\alpha=0$,
we have $\Gamma=c_a^2$ and the bounce is always perfectly
symmetric. This includes bounces with two perfect fluids or scalar
fields dominated by their kinetic energy (compare Refs.
\cite{Fin,PetPin}). The time-reversal symmetry induces several
cancellations in the integrals, allowing the creation of an
intermediate region C, defined by
$(2+c_a^2)/3<c_b^2<min(5/3+2c_a^2,2-c_a^2)$. In this region,
$\tilde{\zeta}_a$ and $\tilde{\zeta}_b$ keep their original
spectrum, without mixing. Nevertheless, $\Psi$ and $\tilde{\zeta}$
are dominated by the first component, so that $\Psi^C\sim
\tilde{\zeta}^C \sim c_1(k)$. Finally, in region B, $\Psi$ and
$\tilde{\zeta}$ get an extra $k^2$ factor w.r.t. the case $\alpha
\neq 0$.

In no case does the spectrum $k^{-2}c_1(k)\sim k^{-\frac{3}{2}-q}$
of the Pre-Bounce growing mode of $\Psi$ survive at the
Post-Bounce horizon re-entry. All the above results have been
confirmed by numerical integration of the equations
\cite{BozVen2}. Our results support the conclusion that a smooth
bounce cannot generate a scale-invariant spectrum via the
mode-mixing mechanism advocated in \cite{Ekp}. One can argue, of
course, that  the way string theory will actually describe the
bounce has no resemblance to a regular bounce described by General
Relativity. In that case, understanding the spectrum of
perturbations in such models will have to wait until a full
stringy treatment of the problem is found. On the other hand,
claims that such a treatment will automatically lead to a flat
spectrum through mode-mixing appear unjustified.

Some authors, working within specific models,  have reached
conclusions that appear to contradict our general analysis. In
Ref. \cite{PetPin}, Eq. (26) provides an analytical solution
$f(\eta)= \eta/(1+\eta^2)^2$ for perturbations outside the horizon
which is valid both far from and through the bounce.
However, the authors use an approximate form
($f_{app}(\eta)=1/\eta^3$) of this solution far from the bounce and
then match $f_{app}$ to $f$ just before the bounce. Had they
used $f$ from the beginning throughout the bounce, they would
have found no mixing. No surprise, therefore, that the mixing they find is at
order $\eta^{-5}$, which is just the next to leading order
term in $f(\eta)$, which is missing in $f_{app}(\eta)$. Had they
 kept the $\eta^{-5}$ term in $f_{app}(\eta)$ they would have
found mixing at the order $\eta^{-7}$. We thus conclude that the
mixing they find is just an artifact of their matching procedure.

In the
case of Ref. \cite{Fin}, the limit $\gamma \rightarrow 0$ ($k
\rightarrow 0$ in our notation), used to evaluate the final
spectrum, just selects the infrared modes at the bounce. However,
the final spectrum should be evaluated at the horizon re-entry,
after the decaying mode has gone away; it would correspond to
$\gamma \rightarrow 0$ while keeping $\gamma z$ (our $k\eta$) of order
1. Finally, a recent paper \cite{CNZ} studying the slow
contraction of the ekpyrotic model in the synchronous gauge agrees
with our general conclusions. However, these authors ignore a
possible divergence of the $\zeta$ variable at the NEC-violation
point, and this appears to restrict significantly the applicability of their
argument.

V.B. thanks the CERN Theory Unit for hospitality. We acknowledge
useful discussions with  F. Finelli, J. Hwang,  V. Mukhanov, P.
Peter and D. Schwarz.

\end{document}